\documentclass[preprint,12pt,3p]{elsarticle}

\usepackage{amssymb}
\usepackage{booktabs}
\usepackage{appendix}
\usepackage{CJK}
\usepackage{stfloats}
\usepackage{amsmath,amssymb,amsfonts}
\usepackage{algorithmic}
\usepackage{graphicx}
\usepackage{textcomp}
\usepackage{xcolor}
\usepackage{algorithm} 
\usepackage{multirow} 
\usepackage{diagbox}
\usepackage{algorithmic}
\usepackage{caption}
\usepackage{makecell}

\journal{Nuclear Physics B}

\begin{document}

\begin{frontmatter}
	
	\title{Analysis and Detection against Network Attacks in the Overlapping Phenomenon of Behavior Attribute}
	
	\address[label2]{Institute of Information Engineering, Chinese Academy of Sciences, Beijing, China}
	\address[label3]{Key Laboratory of Network Assessment Technology, University of Chinese Academy of Sciences, Beijing, China}
	\address[label4]{School of Cyber Security, University of Chinese Academy of Sciences, Beijing, China}
	
	\author[label2,label4]{Jiang Xie}
	\ead{xiejiang@iie.ac.cn}
	
	\author[label2,label3,label4]{Shuhao Li\corref{cor1}}
	\cortext[cor1]{The corresponding author of this paper is Shuhao Li.}
	\ead{lishuhao@iie.ac.cn}
	
	\author[label2,label3,label4]{Yongzheng Zhang}
	\ead{zhangyongzheng@iie.ac.cn}
	
	\author[label2,label4]{Peishuai Sun}
	\ead{sunpeishuai@iie.ac.cn}
	
	\author[label2,label4]{Hongbo Xu}
	\ead{xuhongbo@iie.ac.cn}
	
	\begin{abstract}
		The proliferation of network attacks poses a great threat. Traditional detection methods, whether two-classification or multi-classification, belong to single-label learning and classify a sample into a separate category. However, we discover that there is a noteworthy phenomenon of behavior attribute overlap between attacks in the real world, i.e., a network behavior may be multi-labeled and can be classified into multiple attacks. We verify the phenomenon in well-known datasets(UNSW-NB15, CCCS-CIC-AndMal-2020). In addition, detecting network attacks in a multi-label manner can obtain more information behind them, providing support for tracing the attack source and building IDS. Therefore, we propose a multi-label detection model based on deep learning, MLD-Model, in which WGAN-GP with improved loss performs data enhancement to alleviate the class imbalance problem, and Auto-Encoder(AE) performs classifier parameter pre-training. 
		Experimental results demonstrate that MLD-Model can achieve excellent classification performance. It can achieve $ F1 $=80.06\% in UNSW-NB15 and $ F1 $=83.63\% in CCCS-CIC-AndMal-2020.
		Especially, MLD-Model is 5.99\%$ \sim $7.97\% higher in $ F1 $ compared with the related single-label methods.
		
	\end{abstract}
	
	
\end{frontmatter}

\section{Introduction}
The development of Internet makes us pay more attention to cyber security. It is important to construct corresponding detection schemes for different network attacks and obtain more information from samples\cite{husak2018survey}.
Traditional detection methods just classify attacks into two-classification or multi-classification, which belongs to single-label learning, i.e., a sample has only one label. And there is little related work to explore the correlation of intrinsic features between different attacks.


In this paper, we study the characteristics of various network attacks in the real world, and find that network attacks exist overlapping phenomenon of behavior attribute.
Samples belonging to different network attacks show the same behavior features, are multi-labeled. 
For instance, DoS and Fuzzers attacks will show same features in certain circumstances\cite{moustafa2015unsw}.
The fundamental reason is that malicious behaviors can naturally be defined as different categories from different perspectives. 
We further describe the overlapping phenomenon and analyze the reasons in Section 3 and Section 6.1.

The overlapping phenomenon causes a network behavior may be multi-labeled.
And if we find that a sample belongs to multiple attacks, we can obtain more information from this sample.
Taking DoS and Fuzzers attack in the UNSW-NB15 dataset for example, DoS means a malicious attempt to make a server unavailable to users and Fuzzers mean that attacker attempts to cause a network suspended by feeding it the randomly generated data\cite{moustafa2015unsw}.
Therefore, if a sample is detected as belonging to both DoS and Fuzzers, we can infer that the attacker attempts to make a server unavailable to users by continuously feeding it the randomly generated data. However, traditional methods can only detect this sample as one of DoS or Fuzzers at most, so that we cannot obtain more information (such as the specific method and purpose behind it) to support the tracing of network attack source. 

We call the network attacks exist overlapping phenomenon of behavior attribute as multi-label network attacks.
And it is meaningful to find a multi-label detection method to detect those attacks, which means that we can obtain more information from the detection process to trace the source of network attacks and formulate stronger defense schemes.
However, those detection methods belong to multi-label learning(MLL). 
Currently, they are mainly aimed at natural language processing (nlp) and image fields. 
Nlp mainly includes sentiment classification(\cite{gupta2019distributional,yu2018improving,yilmaz2021multi}), text classification(\cite{zhang2018multi,chang2020taming, roudsari2020multi, banerjee2019hierarchical}), \textit{etc.}, and the image field includes image annotation(\cite{liu2018svm,jing2016multi,wu2015multi}) and image classification(\cite{zha2008joint,zhu2017learning}), \textit{etc.}. 
These detection methods of nlp and image fields cannot be directly used in the processing of network attack data. Currently, there is no multi-label detection technology for network attacks in cyber security. Therefore, it is necessary to find a effective method for the detection of multi-label network attacks.

The contributions of this paper are as follows.

\begin{itemize}
	\item We discover that there is overlapping phenomenon of behavior attribute between network attacks in the real world.
	A network behavior may be multi-labeled and belongs to multiple attacks. We formally describe this phenomenon and analyze its causes.
	\item We perform statistical analysis in well-known network attack datasets (UNSW-NB15\cite{moustafa2015unsw}, CCCS-CIC-AndMal-2020\cite{keyes2021entroplyzer, rahali2020didroid}). The results validate our findings about the overlapping phenomenon. In UNSW-NB15, a sample has 1.689 labels on average.   In CCCS-CIC-AndMal-2020, a sample has 1.413 labels on average. In addition, we process these data and make them publicly available to support related research\footnote{The dataset and code can be found at \textit{https://github.com/BitBrave-Xie/processed-multi-label-dataset}.}. 
	\item We propose a Multi-Label Detection method based on the WGAN-GP\cite{gulrajani2017improved} (with improved loss) and unbalanced Auto Encoder (AE)\cite{baldi2012autoencoders}, MLD-Model, for the detection of multi-label network attacks. WGAN-GP is used for data enhancement to alleviate class imbalance problem. Unbalanced AE is used to extract data features, and pre-training adjust classifier parameters based on the augmented data. Finally, the raw labeled data is used for fine-tuning of the classifier.
	\item We design a prototype system based on MLD-Model, and conduct experiments on two network attack datasets. In UNSW-NB15, there are 10 categories (benign and 9 types of network attacks), and MLD-Model can reach $ Acc $=79.87\%, $ F1 $=80.06\%. In CCCS-CIC-AndMal-2020, there are 15 categories (benign and 14 types of network attacks), and MLD-Model can reach $ Acc $=83.17\%, $ F1 $=83.63\%. Specially, MLD-Model is 5.99\%$ \sim $7.97\% higher in $ F1 $ compared with the related single-label network attack detection methods and 1.65\%$ \sim $58.25\% higher in $ F1 $ compared with other multi-label baseline detection methods.
\end{itemize}



The remainder of this paper is organized as follows. Section 2 introduces the related work. In Section 3, we analyze the overlapping phenomenon of behavior attribute between network attacks. Preliminaries is in Section 4. Subsequently, Section 5 is the methodology and we introduce the composition of MLD-Model. In Section 6, we evaluate our method and show the relevant experimental results. Finally, we discuss and summarize in Section 7 and Section 8, respectively.

\section{Related work}
We analyze and detect various network attacks existing the overlapping phenomenon of behavior attribute, which belong to the field of intrusion detection. The detection method belongs to the field of multi-label learning(MLL). This section will introduce related work from these two perspectives.

\subsection{Intrusion detection}
Intrusion detection against network attacks is one of the important research fields of cyber security. An Intrusion Detection System (IDS)\cite{liao2013intrusion} is built based on various technical means to detect, resist and warn against network attacks. 
Existing methods for intrusion detection can generally be divided into feature detection and anomaly detection.
Feature detection, also called misuse detection, fits the behavior patterns of known attacks, and judges network behaviors with similar behavior patterns as malicious.
Feature detection can maintain a high detection rate for known network attacks,
but it cannot effectively detect unknown attack, i.e., the Zeroday attack.
Anomaly detection, also called behavior detection, is one of the mainstream methods of intrusion detection. It mainly fits the patterns of normal network behavior. When a network behavior does not conform to the pattern of the feature library, it is judged as malicious.
Therefore, anomaly detection 
can detect Zeroday attack more effectively, which is important for the current cyber security situation where new network attacks continue to emerge.


Zhiqiang \textit{et al.} \cite{Zhiqiang2019modeling} propose a IDS based on deep learning, and experiments show that the proposed classifier is better than other models.
Kumar \textit{et al.} \cite{kumar2020integrated} propose a IDS based on feature detection, which can detect 5 types of intrusions in the network: Exploits, DoS, Probe, Generic, and Normal.
Yang \textit{et al.} \cite{yang2019improving} propose a novel intrusion detection model called ICVAE-DNN. The NSL-KDD\cite{tavallaee2009detailed} and UNSW-NB15 datasets are used to evaluate. Experiments show that it outperforms 6 well-known models. Then, Yang \textit{et al.}\cite{yang2020network} also propose a network intrusion detection model called SAVAER-DNN, which uses WGAN-GP to learn the latent data distribution. Experiments show that the SAVAER-DNN outperforms 8 well-known classification models.
Jing \textit{et al.} \cite{Jing2019svm} propose a SVM for two-classification and multi-classification. Experiments in UNSW-NB15 show that the proposed method can achieve accuracy of 75.77\%.

Durmucs \textit{et al.} \cite{Durmucs2021analysis} apply statistical calculation methods to the analysis output model. Then, an understandable scenario is created and a model for cyber security intervention is provided.
Fiky \textit{et al.} \cite{Fiky2021android} proposes two machine learning methods for dynamic analysis of Android malware: one is used to detect and identify the Android malware category, and the other is used to detect and identify the Android malware family. In general, a method for high-precision dynamic analysis of Android malware is provided, and an accuracy rate of more than 96\% can be obtained in the CCCS-CIC-AndMal2020 dataset.
Liu \textit{et al.} \cite{Liu2021research} applies an unsupervised malware detection method in order to detect Zeroday attack, and proposes an unsupervised feature learning algorithm called Restricted Boltzmann Machine based on Subspace (SRBM) \cite{le2008representational} to reduce the data dimension. Experimental results show that the features learned by SRBM perform better than those learned by other feature reduction methods.
Abusitta \textit{et al.} \cite{Abusitta2021robust} proposes a new framework for detecting malware in a non-stationary environment. It uses deep learning technology to extract useful features and is robust to changing environments. Experimental results on the actual dataset show that the framework improves the detection accuracy compared with the existing methods.

The above-mentioned various detection methods have excellent performance on the single-label classification of network attacks. However, the overlapping phenomenon of behavior attribute between network attacks is not considered.
This results in their theoretical upper limit of accuracy not being 100\%.

\subsection{Multi-label learning}
Multi-label learning(MLL) tasks are more difficult than single-label tasks. Multi-label means that a sample has multiple labels.
Multi-label learning(MLL) classification tasks are more difficult than single-label classification tasks. Multi-label means that a sample has multiple labels. In some cases, these labels have priority.
The difficulty in constructing a multi-label learning algorithm is mainly due to the exponential growth of the output space. For instance, in the multi-label learning of $ M $ basic categories, the theoretical output space is $ 2^{M} $.
According to the strength of label correlation mining, methods for multi-label detection mainly have three strategies: first-order, second-order, and high-order\cite{zhang2013review}.

	\textbf{First-order:} It ignores the correlation with labels, and only builds binary-classifiers between single-label and 	samples\cite{boutell2004learning, zhang2007ml, clare2001knowledge}.
    For instance, a multi-label classification of $ M $ basic categories, is decomposed into $ M $ independent 2-class problems. 
	First-order is simple and can be quickly constructed using basic classifiers.
	However, the correlation between labels cannot be effectively used in this way.

\textbf{Second-order:} 
It explores the correlation features between pairs of labels, such as divides the labels into related and unrelated sets\cite{ghamrawi2005collective, qi2007correlative, ueda2003parametric}, or divide the labels into related and unrelated sets\cite{elisseeff2001kernel, furnkranz2008multilabel}.
For instance, in a multi-label problem of $ M $ basic categories, it constructs $\frac{M(M-1)}{2}$ two-classifiers of label pairs. Compared with the first-order, the second-order considers the correlation between label pairs.

\textbf{High-order:} It considers the association between multiple labels\cite{cheng2009combining, godbole2004discriminative, ji2008extracting, yan2007model}.
For instance, the label subset is directly converted into a specific natural number. It converts the multi-label problem into a single-label multi-classification problem. 
Generally, high-order can achieve the better detection results, 
but the structure is also relatively more complicated.

There are two mainstream methods based on the above three strategies: problem transformation and algorithm adaptation.
Problem transformation is to convert multi-label into a single-label combination of two-classification problems or multi-classification problems, and then use the corresponding mature algorithms. Algorithm adaptation is to directly adopt existing algorithms, such as ML-KNN\cite{zhang2007ml} determines the label subset of samples based on the neighbor features.

Many researchers conduct multi-label detection in cyber security. Li \textit{et al.}\cite{li2019extraction} conduct apt-related potential threat detection. 
Han \textit{et al.}\cite{han2018multi} proposes a weakly supervised multi-label learning method based on the collaborative embedding to solve the problem of incomplete data collection. 
However, there is no relevant research on multi-label learning algorithms for network attack detection in the overlapping phenomenon of behavior attribute. 


\section{Analysis of the overlapping behavior attribute}



\begin{figure}[htbp]
	\centerline{\includegraphics[scale=0.27]{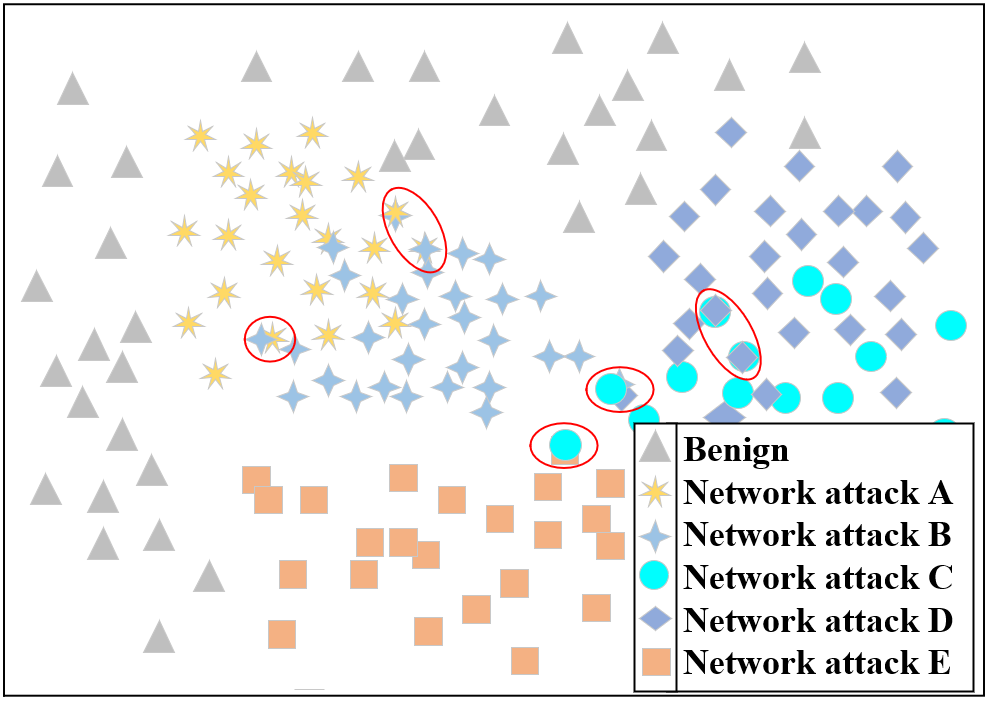}}
	\caption{The overlapping phenomenon of behavior attribute between network attacks.}
	\label{abcdef}
\end{figure}

We discover that there is the overlapping phenomenon of behavior attribute between network attacks in the real world.
Various network attacks are not clear-cut, but overlap and contain each other. 
We show a demo in Fig.\ref{abcdef}, assuming that network attacks are distributed in a two-dimensional space. As shown in the Fig.\ref{abcdef}, each point represents a sample, and it can be seen that network attacks A and B have overlap of behavior attribute, and some samples of network attacks C and D also partially overlap of behavior attribute. 
The phenomenon causes a sample may be multi-labeled. Here we cite some typical cases in practice, and give an analysis. The specific data analysis is in Section 6.1. 

\subsection{Definition of the overlapping behavior attribute} 
The formal description of the overlapping phenomenon is as follows. In a network attack dataset $ D$, a sample $ x=[x^{(1)},x^{(2)},...,x^{(d)}] $.
$ D $ consists of multiple attack sub-datasets ($ D_1,D_2,...,D_M $), where the attack $ i $ has $ \left| D_i\right| $ samples.
We define two samples $ x $ and $ x' $ with overlapping behavior attribute as $ x=x' $, which are exactly the same in all features, as follows:

	\begin{equation*}
		\begin{aligned}
			x=x' \Leftrightarrow  \left\{ x^{(i)}=x'^{(i)}; i=1,2,...,d \right\}
		\end{aligned}\label{eqx}
	\end{equation*}

Therefore, we define that if there is the overlapping phenomenon of behavior attribute between network attacks in $ D $, then: $ \exists \left ( x_1\in D_1, x_2\in D_2, ..., x_k\in D_k; k\leqslant M \right )$, has $x_1=x_2=...=x_k$.

\subsection{Case analysis of the overlapping behavior attribute} 

\textbf{Analysis, Backdoor, DoS, Exploits and Fuzzers in UNSW-NB15:}
There are a total of 9 attacks in UNSW-NB15. And there is the overlapping phenomenon
of behavior attribute in the Analysis, Backdoor, DoS, Exploits and Fuzzers, i.e., they have same records. Taking DoS and Fuzzers for example. DoS is a malicious attempt to make a server unavailable to users and Fuzzers are a technique that attempts to cause a program or network suspended by feeding it the randomly generated data \cite{moustafa2015unsw}. Therefore, if an attacker uses the technology of Fuzzers during the DoS attack, then this behavior can be considered to belong to both DoS and Fuzzers.

\textbf{Trojan and Zeroday in CCCS-CIC-AndMal-2020:} 
Trojan is a software or script that accepts instructions to perform malicious actions in the victim's host\cite{rahali2020didroid}. The attacker usually communicate with the Trojan by C\&C channels. Zeroday is relatively broader. Generally, any network attack that uses unknown vulnerabilities or backdoors can be considered Zeroday\cite{keyes2021entroplyzer, rahali2020didroid}.
Therefore, if an attacker uses unknown vulnerabilities to transmit Trojan or directly uses it as a C\&C channel, it can be considered to belong to both Trojan an Zeroday.

There are other overlapping phenomenon, such as Reconnaissance and Exploits in UNSW-NB15. In general, the overlapping phenomenon between network attacks is an universal problem that cannot be ignored.


\subsection{Cause analysis of the overlapping behavior attribute}
We analyze why there is the overlapping behavior attribute between network attacks. And we believe that there are the following reasons based on experience and survey results.
\begin{itemize}
	\item The conceptual definitions of different attacks contain each other, so that a network behavior is inherently multi-labeled. In some circumstances, the definition of two attacks can be considered the same because the attack methods used are the same (such as the Trojan and Zeroday). 
	\item The actions taken by the attacker in the process of implementing the attack behavior are complex, and the malicious features exposed from different stages make a same behavior be classified into different attack types.
	\item The existing feature extraction methods are incompleteness, so that the unique mutually exclusive features of attack are not extracted. For instance, although the external features are very similar for two different attacks based on the same C\&C encrypted channel, the content that mainly show the characteristics of the attack cannot be directly represented by those statistics features.
\end{itemize}

In this paper, we investigate the overlapping phenomenon of behavior attribute between network attacks in UNSW-NB15 and CCCS-CIC-AndMal-2020.
And we quantitatively verify the phenomenon in Section 6.1.

\section{Preliminaries}
The overlapping phenomenon enables the network attack samples to be multi-labeled. Therefore, we construct a multi-label detection method based on WGAN-GP and AE.
In this section, we first introduce the multi-label classification problem. Then the WGAN-GP and AE are introduced.

\subsection{Definition of multi-label attack classification problem}
We consider network attack detection problem as multi-label learning based on the overlapping phenomenon of behavior attribute. 
There is $ \mathbf{X} \in \mathbb{R}^{d} $, $ d $ is the size of feature space, $ \mathbf{Y} = \left \{ y_{1},y_{2},...,y_{M} \right \} $ means the label space. There is a network attack dataset $ \mathbf{D} = \left \{ \left ( x_{i},Y_{i} \right ) | i=1,2,...,N; x_{i}\in \mathbf{X};Y_{i} \subseteq \mathbf{Y} \right \} $. Our task is to find a multi-label classification function $h$ that maps $x$ from the feature space to the label space. For a sample $x$, $h(x) \subseteq \mathbf{Y}$ is given as the label set of the sample.

\subsection{WGAN-GP}
\begin{figure}[htbp]
	\centerline{\includegraphics[scale=0.45]{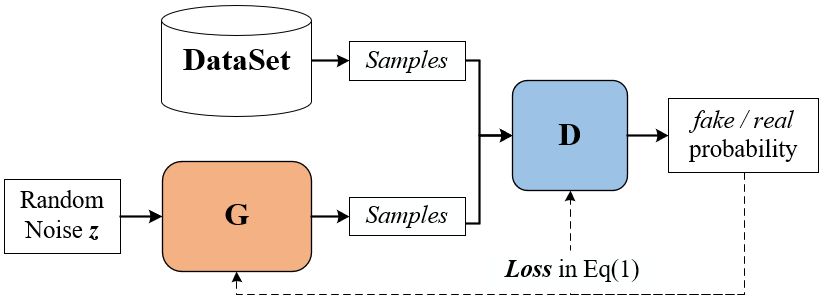}}
	\caption{The infrastructure of traditional WGAN-GP.}
	\label{WGAN-GP}
\end{figure}

WGAN-GP\cite{gulrajani2017improved} is a neural network model. 
It provides a powerful algorithm framework for unsupervised learning. As shown in Fig.\ref{WGAN-GP}, WGAN-GP includes generator $G$ and discriminator $D$. $G$ is used to imitate the real data distribution $\mathbb{P}_{r}$ to generate a fake data distribution $\mathbb{P}_{g}$. $D$ is used to determine whether the sample is generated by $G$.
$G$ and $D$ play against each other and finally strike a balance.

WGAN-GP is more stable and easier to converge than vanilla GAN, and can generate more diverse data.
Vanilla GAN adopts JS divergence to measure the difference between $ \mathbb{P}_{r} $ and $ \mathbb{P}_{g} $. However, the JS divergence is infinite when the two distributions have no intersection, so it is easy to have gradients disappear.
WGAN adopts Earth-Mover distance instead, referred to as EM distance, or Wasserstein distance to calculate the difference between two distributions.
However, tThe 1-Lipschitz constraint need to be satisfied in EM distance. Therefore, the gradient penalty (GP) is added to the loss of WGAN-GP, as shown in Eq\eqref{eq1}, where $ \mathbb{P}_{\hat{x}} $ represents the distribution of all data and $ \lambda$ is the weight.

	\begin{equation}
		\begin{aligned}
			\mathbf{L} = \underbrace{\underset{\tilde{x} \sim \mathbb{P}_{g}}{\mathbb{E}}[D(\tilde{x})]-\underset{x \sim \mathbb{P}_{r}}{\mathbb{E}}[D(x)]}_{\text {Original loss }}+\underbrace{\lambda \underset{\hat{x} \sim \mathbb{P}_{\hat{x}}}{\mathbb{E}}\left[\left(\left\|\nabla_{\hat{x}} D(\hat{x})\right\|_{2}-1\right)^{2}\right]}_{\text {Gradient penalty }}\\ 
		\end{aligned}\label{eq1}
	\end{equation}


\subsection{Auto-Encoder}
Auto-Encoder (AE)\cite{baldi2012autoencoders} is a feed-forward neural network, usually used for data dimension reduction and feature extraction. Different from the traditional neural network focusing on the final loss reduction and the improvement of classification accuracy, AE focuses on extracting effective features from the data and reconstructing them. 

\begin{figure}[htbp]
	\centerline{\includegraphics[scale=0.35]{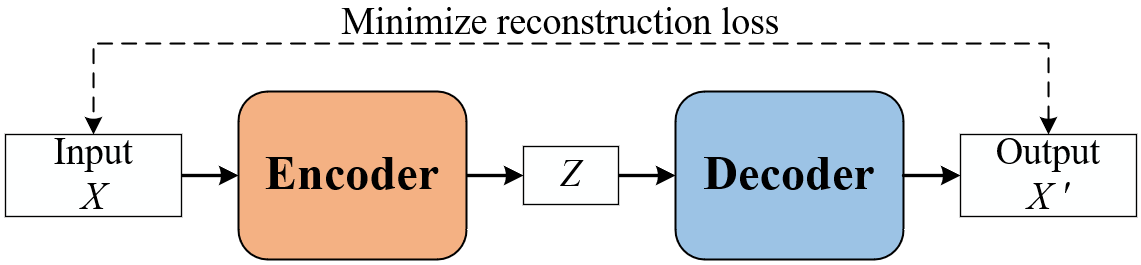}}
	\caption{The infrastructure of traditional AE.}
	\label{AE}
\end{figure}

AE is mainly composed of two parts, encoder and decoder, as shown in Fig.\ref{AE}. Encoder is used to map the features of raw data to another feature space. Decoder reconstructs the converted features back to the raw space to ensure that it better inherits the feature information of the raw data.

\section{Methodology} 
\subsection{Overview}
In this paper, we build MLD-Model to detect multi-label network attack in overlapping phenomenon.
First, WGAN-GP with improved loss is used for data enhancement to alleviate the class imbalance problem.
Then, unbalanced AE performs unsupervised pre-training.
Finally, the pre-trained encoder and a softmax classification layer are combined together as a neural network classifier, and the raw labeled data is used for parameter fine-tuning.
We adopts the Label PowerSet scheme\cite{tai2012multilabel} for multi-label learning. Each label subset is mapped to a natural number, and the multi-label problem is converted to a single-label multi-classification problem.

\begin{figure}[htbp]
	\centerline{\includegraphics[scale=0.47]{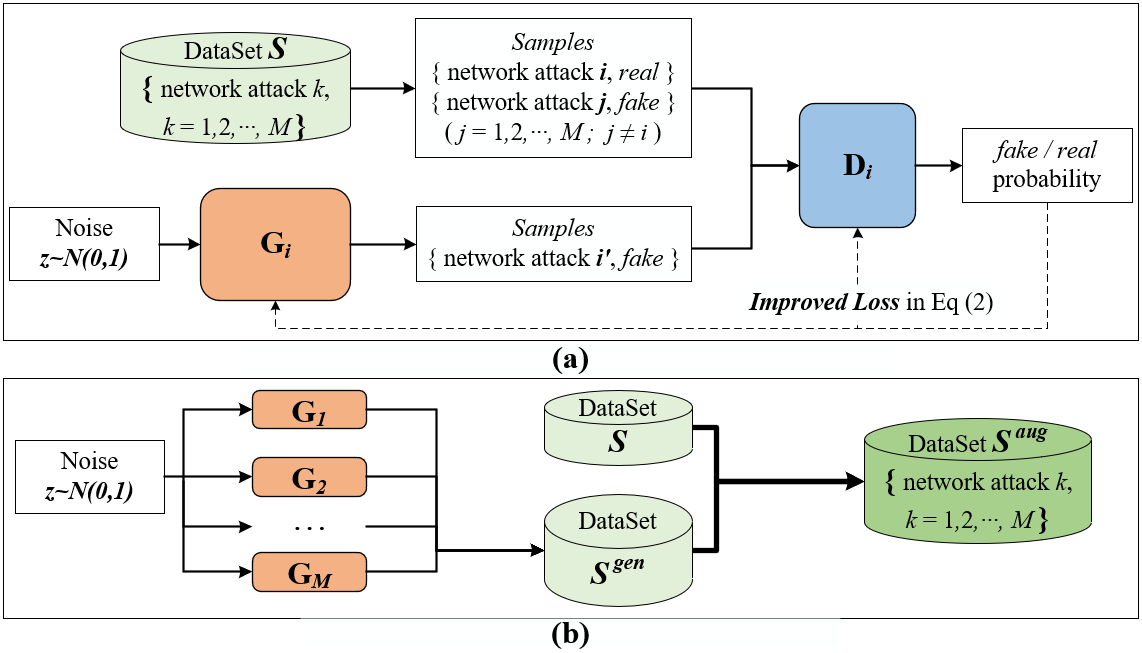}}
	\caption{The process of data enhancement based on the WGAN-GP with improved loss in MLD-Model. (a) Training WGAN-GP$ _i $  with improved loss in Eq\eqref{eq2}. (b) Generating network attack data. ($ i $: category of network attack ($ i=1,2,...,M $); $ S $: the raw dataset; $ S^{gen} $: the generated dataset; $ S^{aug} $: the augmented dataset).}
	\label{WGAN-GPxj}
\end{figure}

\subsection{Data enhancement}
The overall process of data enhancement based on the WGAN-GP with improved loss is shown in Fig.\ref{WGAN-GPxj}.
In the training phase(Fig.\ref{WGAN-GPxj}(a)), when a network attack needs to be generated, the corresponding samples are marked as $ real $, and other attack data is marked as $ fake $.
In the generation phase(Fig.\ref{WGAN-GPxj}(b)), noise is input to $G$ to obtain generated samples.
Finally, an augmented dataset $ S^{aug} $ consisting of the raw dataset $ S $ and the generated dataset $ S^{gen} $ is constructed.

In this process, we adopt WGAN-GP with improved loss to generate fake attack data.
Traditional WGAN-GP only considers that the generated data should be in the same space as the corresponding real data.
In the overlapping phenomenon, however, we need to consider that the generated data should also be different from other attacks. Otherwise, the generated data will increase the clutter of feature space.

	\begin{equation}
		\begin{aligned}
			\mathbf{L} &= \underbrace{\underset{\tilde{x} \sim \mathbb{P}_{g}}{\mathbb{E}}[D(\tilde{x})]-\underset{x \sim \mathbb{P}_{r_{i}}}{\mathbb{E}}[D(x)]}_{\text {Original loss }}+\underbrace{\lambda \underset{\hat{x} \sim \mathbb{P}_{\hat{x}_{}}}{\mathbb{E}}\left[\left(\left\|\nabla_{\hat{x}} D(\hat{x})\right\|_{2}-1\right)^{2}\right]}_{\text {Gradient penalty }}\\
			&+\  \underbrace{\lambda' \underset{x' \sim \mathbb{P}_{r_{j}(i\neq j; j={1,2,...,M})}}{\mathbb{E}}[D(x')]}_{\text {Our penalty item }}
		\end{aligned}\label{eq2}
	\end{equation}

Therefore, we update the traditional loss(Eq\eqref{eq1}) to improved loss(Eq\eqref{eq2}). As shown in the Eq\eqref{eq2}, $ \mathbb{P}_{r_{i}} $ is the data distribution of attack $ i $, $ \mathbb{P}_{r_{j}(i\neq j; j={1,2,...,M})} $ is the data distribution of other attacks except $ i $ and $ \lambda' $ is the weight.
A penalty item is added when the neural network generates fake data of attack $ i $. WGAN-GP with improved loss will generate the corresponding attack data while maintaining the difference between it and other attacks as much as possible.

In addition, the generated data is only used in the pre-training phase.
Because feature space of network attacks in the overlapping phenomenon are nested with each other, WGAN-GP can only partially simulate the distribution of attacks. 
In experiments, the generated data will interfere with the decision of the classifier if it is used for model fine-tuning.

\begin{figure}[htbp]
	\centerline{\includegraphics[scale=0.350]{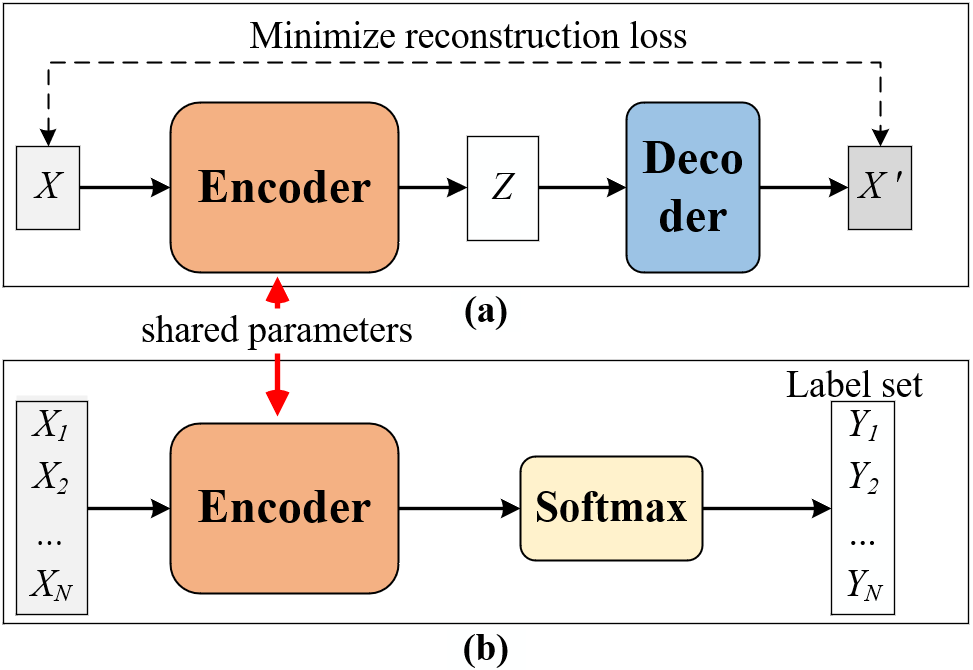}}
	\caption{The pre-training phase and classification phase in MLD-Model. (a) Pre-training phase. (b) Classification phase. ($ X $: the data in the augmented dataset $ S^{aug} $; $ X_i $: the data in raw dataset $ D $ with multi-label; $ Y_i $: the label set of $ X_i $ detected by the classifier).}
	\label{AE-XJ}
\end{figure}

\subsection{Pre-training}
We call the AE used for pre-training as unbalanced AE, because the network structure of encoder and decoder are asymmetrical.
The encoder has more neurons so that more feature extraction operations can be retained in it.
In addition, data in $ S^{aug} $ is used for unsupervised pre-training of the unbalance AE. Experiments show that the model can converge faster and achieve better results after the parameters of the neural network are pre-adjusted.
The details are shown in the Pre-training phase(Fig.\ref{AE-XJ}(a)).

\subsection{Classification}
The classifier is a neural network formed by stacking the encoder in the pre-trained unbalanced AE and a softmax layer. Since the main parameters of classifier have been pre-trained, we only need to use the raw data for fine-tuning. The specific process is shown in the classifier phase in Fig.\ref{AE-XJ}(b).

\subsection{Algorithm}
The overall training and detection process of MLD-Model is shown in Algorithm 1. The input is network attack data and hyper-parameters, and the output is the corresponding unknown network attack multi-label set.
In the data preprocessing phase, we analyze the network attack dataset $ S $ to obtain the multi-label attack dataset $ D $. 
In the data enhancement phase, WGAN-GP with improved loss is used to construct the network attack generated dataset $ S^{gen} $, and then combined with $ S $ to form the augmented dataset $ S^{aug} $. 
In the pre-training phase, unsupervised pre-training based on the unbalanced AE is performed using the data in $ S^{aug} $.

In the classifier phase, the multi-label attack data in $ D $ is used to fine-tune the model parameters.
In the final detection phase, the classifier judges the attack multi-label set of each sample in the unknown network attack dataset $ D' $, and gives the multi-label detect results of unknown network attacks.

\begin{algorithm} 
	\caption{The training and detection process of MLD-Model} 
	\label{alg2} 
	\begin{algorithmic}[1] 
		\REQUIRE network attack dataset $ \mathbf{S} = \left \{ \left ( x_{i},y_{i} \right ) | i=1,2,...,N; x_{i}\in \mathbf{X};y_{i} \in \mathbf{Y} \right \} $; unknown network attack dataset $ \mathbf{D}' = \left \{ \left ( {x'}_{i} \right ) | i=1,2,...,N'; x'_{i}\in \mathbf{X} \right \} $; Hyper-parameters $ P $: Various hyper-parameters (number of neurons, etc.) needed to construct MLD-Model.
		
		\ENSURE $ \mathbf{Y}' $: Multi-label results of unknown network attacks.
		\STATE{\textbf{Step 0:} Data preprocessing}
		\STATE{network attack multi-label dataset $ D = \{\} $}
		\FOR { $ (x_{i},y_{i}) $ in $ S $} 
		\IF {$ x_i $ in D} 
		\STATE {$ D = (D - \{(x_i,Y_i)\}) \cup  \{ (x_i,Y_i \cup \{y_i\}) \} $ }
		\ELSE 
		\STATE {$ D = D \cup \{(x_i,Y_i)\} $}
		\ENDIF
		\ENDFOR
		
		\STATE{\textbf{Step 1:} Data enhancement}
		\STATE{build the generated dataset $ S^{gen} = \{\}$}
		\FOR { network attack $ q $ in $ S $} 
		\STATE {$ S_q = \{(x_i,real) | i=1,2,...,N; y_i=q\} $ }
		\STATE {$ \bar{S}_q = \{(x_i,fake) | i=1,2,...,N; y_i\neq q\} $ }
		\STATE {WGAN-GP$_i \leftarrow  $ trained WGAN-GP with improved loss (Eq(2)) based on dataset $ S_i $ and $ \bar{S}_q $}
		\STATE {$ S^{gen} = S^{gen} \cup $ $ \{ $ data generated by WGAN-GP$_i\} $ }
		\ENDFOR
		\STATE {build the augmented dataset $ S^{aug} = S^{gen} \cup S $}
		
		\STATE{\textbf{Step 2:} Pre-training}
		\STATE {unbalanced AE = encoder + decoder}
		\STATE {unbalanced AE $ \leftarrow  $ trained unbalanced AE based on $ S^{aug} $}
		
		\STATE{\textbf{Step 3:} Classification}
		\STATE {classifier = encoder(in unbalanced AE) + softmax}
		\STATE {classifier $ \leftarrow  $ trained classifier based on $ D $}
		
		\STATE{\textbf{Step 4:} Detection}
		\STATE {$ Y'=\{\} $}
		\FOR { $ x'_i $ in $ D' $} 
		\STATE {$ Y'_i = $ classifier($ x'_i $) $ \subseteq \mathbf{Y} $ }
		\STATE {$ Y' = Y' \cup \{(x'_i,Y'_i)\} $ }
		
		\ENDFOR
		
		\RETURN $ Y' $
	\end{algorithmic} 
\end{algorithm}

\subsection{Summary}
MLD-Model focuses on the detection of multi-label network attacks. 
The data enhancement and pre-training techniques are used in the training process to improve the detection performance.

In the data enhancement phase, we adopt WGAN-GP as the basic architecture. Because more stable than vanilla GAN, WGAN-GP can better resist gradient disappearance/explosion, generate more diverse and real network attack data. In addition, we propose the improved loss (Eq(2)) to replace the traditional loss. 

In the pre-training phase, we set the structure of the encoder and decoder in AE to be asymmetric. There are more neurons in the encoder to obtain stronger feature extraction capabilities. 
Finally, the classifier combined with the pre-trained encoder can be fine-tuned based on the processed raw multi-labeled dataset to obtain excellent performance.


\section{Experiment and evaluation}

We verify the overlapping phenomenon of behavior attribute between network attacks in UNSW-NB15 and CCCS-CIC-AndMal-2020. Then, we sample data from those two datasets for multi-label detection of MLD-Model.

\subsection{Data analysis of the overlapping behavior attribute}

\subsubsection{Data collection and sampling}
\textbf{UNSW-NB15}\cite{moustafa2015unsw}, is a dataset that hybrids the real modern normal and the contemporary synthesized attack activities of the network traffic. The major categories of the records are normal and attack. The attack records are further classified into 9 families. The raw network packets of the UNSW-NB15 dataset is created by the IXIA PerfectStorm tool in the Cyber Range Lab of the Australian Centre for Cyber Security (ACCS). The number of records in the training set is 175,341 and in the testing set is 82,332. 
\textbf{CCCS-CIC-AndMal-2020}\cite{keyes2021entroplyzer, rahali2020didroid}, is a new comprehensive and huge malware dataset from the Canadian Institute for Cybersecurity (CIC) project in collaboration with Canadian Centre for Cyber Security (CCCS). It includes 200K benign and 200K malware samples totaling 400K android apps with 14 prominent malware categories and 191 eminent malware families. These malware mainly exhibits malicious characteristics through network behavior. We show the data sampling results in Tab.\ref{nb15} and Tab.\ref{cicandmal}. 

\begin{table}[htbp]
	\caption{The data sampling result of UNSW-NB15}
	\vspace{-1.5em}
	\begin{center}
		\begin{tabular}{|c|c||c|c|}
			\hline
			\textbf{Category} & \makecell*[c]{\textbf{Size} \\ (Training set + Test set)} & \textbf{Category} & \makecell*[c]{\textbf{Size} \\ (Training set + Test set)} \\
			\hline
			\hline
			Normal & \makecell[c]{93,000 (56,000+37,000)} & Reconnaissance & \makecell[c]{13,987 (10,491+3,496)} \\
			Backdoor & \makecell[c]{2,329 (1,746+583)} & Exploits & \makecell[c]{44,525 (33,393+11,132)} \\
			Analysis & \makecell[c]{2,677 (2,000+677)} & DoS & \makecell[c]{16,353 (12,264+4,089)} \\
			Fuzzers & \makecell[c]{24,246 (18,184+6,062)} & Worms & \makecell[c]{174 (130+44)} \\
			Shellcode & \makecell[c]{1,511 (1,133+378)} & Generic & \makecell[c]{58,871 (40,000+18,871)} \\
			\hline
			\hline
			\textbf{Total} & \makecell*[c]{257,673 (175,341+82,332)} & \multicolumn{2}{l|}{}\\
			\hline
		\end{tabular}
		\label{nb15}
	\end{center}
	\vspace{-1.5em}
\end{table}

\begin{table}[htbp]
	\caption{The data sampling result of CCCS-CIC-AndMal-2020}
	\vspace{-1.5em}
	\begin{center}
		\begin{tabular}{|c|c||c|c||c|c|}
			\hline
			\textbf{Category} & \makecell*[c]{\textbf{Size}} & \textbf{Category} & \makecell[c]{\textbf{Size}} & \textbf{Category} & \makecell[c]{\textbf{Size}} \\
			\hline
			\hline
			Riskware & \makecell[c]{97,349} & Adware & \makecell[c]{47,198} & FileInfector & \makecell[c]{669}\\
			Benign & \makecell[c]{32,084} & Trojan & \makecell[c]{13,542} &  Backdoor & \makecell[c]{1,538} \\
			Zeroday & \makecell[c]{13,327} & Ransomware & \makecell[c]{6,202} & Banker & \makecell[c]{887} \\
			Spy & \makecell[c]{3,540} & SMS & \makecell[c]{3,125} & PUA & \makecell[c]{2,051} \\
			Dropper & \makecell[c]{2,302} & NoCategory & \makecell[c]{2,296} & Scareware & \makecell[c]{1,556} \\
			\hline
			\hline
			\textbf{Total} & \makecell*[c]{227,666} & \multicolumn{4}{l|}{}\\
			\hline
		\end{tabular}
		\label{cicandmal}
	\end{center}
\end{table}

For UNSW-NB15, we select the officially training set and test set, as shown in Tab.\ref{nb15}. A sample has 42 features. A detailed description of these attacks can be found here\cite{moustafa2015unsw}.
For CCCS-CIC-AndMal-2020, the data is randomly sampled.
Since there are too many benign data, we only select part of the benign data (Ben0.csv) and all the malicious data. The sampling results are shown in Tab.\ref{cicandmal}. The sampling results are shown in Tab.\ref{cicandmal}. A sample has 9,503 features. A detailed description of these attacks can be found here\cite{keyes2021entroplyzer, rahali2020didroid}.

\subsubsection{Analysis results}

In the overlapping phenomenon of behavior attribute between network attacks, some samples are multi-labeled. 
We show the specific overlap of attribute based on the data sampled from UNSW-NB15 and CCCS-CIC-AndMal-2020.

The results in UNSW-NB15 are shown in Tab.\ref{unswnb15mul}. For instance, the number Total=188 in the $ sample\ 1 $ column means that there are 188 samples that are the same as $ sample\ 1  $, 66 of which are labeled as DoS and 76 are labeled as Exploits.

\begin{table}[htbp]
	\caption{The top 5 multi-label network attack samples in the overlapping phenomenon of behavior attribute in UNSW-NB15}
	\vspace{-1.5em}
	\begin{center}
		\setlength{\tabcolsep}{0.9mm}{
			\begin{tabular}{|c|c|c|c|c|c|c|}
				\hline
				\multirow{2}{*}{\textbf{Category}} & \multicolumn{6}{c|}{\makecell*[c]{\textbf{Network attack records}}} \\
				\cline{2-7} 
				& \makecell[c]{sample 1} & {sample 2} & {sample 3} & {sample 4} & {sample 5} & \makecell[c]{...}\\
				\hline
				\hline
				\makecell[c]{Analysis} & {10} & {7} & {6} & {6} & {5} & \\
				\makecell[c]{Backdoor} & {10} & {7} & {4} & {6} & {4} & \\
				\makecell[c]{DoS} & {66} & {47} & {38} & {42} & {31} & \\
				\makecell[c]{Exploits} & {76} & {62} & {60} & {54} & {50} & \\
				\makecell[c]{Fuzzers} & {10} & {7} & {6} & {6} & {5} & \\
				\makecell[c]{Generic} & {6} & {1} & {0} & {0} & {0} & {...}\\
				\makecell[c]{Normal} & {0} & {0} & {0} & {0} & {0} & \\
				\makecell[c]{Reconnaissance} & {10} & {7} & {6} & {6} & {5} & \\
				\makecell[c]{Shellcode} & {0} & {0} & {0} & {0} & {0} & \\
				\makecell[c]{Worms} & {0} & {0} & {0} & {0} & {0} & \\
				\hline
				\hline
				\makecell[c]{\textbf{Total}} & {188} & {138} & {120} & {120} & {100} & ... \\
				\hline
		\end{tabular}}
		\label{unswnb15mul}
	\end{center}
\end{table}

The results in CCCS-CIC-AndMal-2020 are shown in Tab.\ref{cicandmalmul}. For instance, the number Total=5,975 in the $ sample\ 1 $ column means that there are 5,975 samples that are the same as $ sample\ 1  $, 3,612 of which are labeled as Trojan and 2,363 are labeled as Zeroday.

\begin{table}[htbp]
	\vspace{-0.5em}
	\caption{The top 5 multi-label network attack samples in the overlapping phenomenon of behavior attribute in CCCS-CIC-AndMal-2020}
	\vspace{-1.5em}
	\begin{center}
		\setlength{\tabcolsep}{0.9mm}{
			\begin{tabular}{|c|c|c|c|c|c|c|}
				\hline
				\multirow{2}{*}{\textbf{Category}} & \multicolumn{6}{c|}{\makecell*[c]{\textbf{Network attack records}}} \\
				\cline{2-7} 
				& \makecell[c]{sample 1} & {sample 2} & {sample 3} & {sample 4} & {sample 5} & \makecell*[c]{...}\\
				
				\hline
				\hline
				\makecell[c]{Adware} & {0} & {206} & {3} & {134} & {1,547} & \\
				\makecell[c]{Backdoor} & {0} & {0} & {0} & {58} & {0} & \\
				\makecell[c]{Banker} & {0} & {0} & {0} & {28} & {0} & \\
				\makecell[c]{Benign} & {0} & {0} & {0} & {12} & {0} & \\
				\makecell[c]{Dropper} & {0} & {0} & {0} & {94} & {0} & \\
				\makecell[c]{FileInfector} & {0} & {0} & {0} & {5} & {0} & \\
				\makecell[c]{NoCategory} & {0} & {0} & {0} & {59} & {0} & \\
				\makecell[c]{PUA} & {0} & {0} & {0} & {7} & {0} & ...\\
				\makecell[c]{Ransomware} & {0} & {0} & {0} & {1,057} & {0} & \\
				\makecell[c]{Riskware} & {0} & {2,602} & {2,544} & {288} & {186} & \\
				\makecell[c]{SMS} & {0} & {0} & {0} & {14} & {0} & \\
				\makecell[c]{Scareware} & {0} & {0} & {0} & {8} & {0} & \\
				\makecell[c]{Spy} & {0} & {0} & {0} & {199} & {0} & \\
				\makecell[c]{Trojan} & {3,612} & {0} & {0} & {93} & {0} & \\
				\makecell[c]{Zeroday} & {2,363} & {0} & {0} & {301} & {0} & \\
				\hline
				\hline
				\makecell[c]{\textbf{Total}} & {5,975} & {2,808} & {2,547} & {2,357} & {1,733} & ...\\
				\hline
		\end{tabular}}
		\label{cicandmalmul}
	\end{center}
	\vspace{-1.5em}
\end{table}

In addition, multi-label metrics are also used to show the overlapping phenomenon, so as to quantitatively measure the overlapping distribution.
The label diversity, $ LDiv $, is the number of different label sets in the dataset, as shown in the Eq\eqref{eq3}.
The label cardinality, $ LCard $, is the average label number of one sample, as shown in the Eq\eqref{eq4}.
Two metrics are calculated to measure the multi-label distribution of network attacks.

\begin{equation}
\begin{aligned}
\operatorname{LDiv} =\left| \left\{ Y\mid\exists \boldsymbol{x}:(\boldsymbol{x}, Y)  \in \mathbf{D} \right\}  \right|
\end{aligned}\label{eq3}
\end{equation}

\begin{equation}
\begin{aligned}
\operatorname{LCard} =\frac{1}{N} \sum_{i=1}^{N} \mid Y_{i}\mid
\end{aligned}\label{eq4}
\end{equation}

The data analysis results about multi-label are shown in Tab.\ref{data-muti}. In UNSW-NB15, there are a total of 10 basic categories and a total of 57 label sets. On average, each sample belongs to 1.689 categories.
In CCCS-CIC-AndMal-2020, there are a total of 15 basic categories, and a total of 145 label sets. On average, each sample belongs to 1.413 categories.
All processed data can be obtained from here\footnote{https://github.com/BitBrave-Xie/processed-multi-label-dataset}. 

\begin{table}[htbp]
	\caption{The data analysis results of network attack overlap}
	\vspace{-1.5em}
	\begin{center}
		\begin{tabular}{|c|c|c|c|}
			\hline
			& \textbf{Basic Category} & \makecell*[c]{\textbf{LDiv}} & \textbf{LCard} \\
			\hline
			\hline
			UNSW-NB15 & \makecell[c]{10} & 57 & 1.689 \\
			CCCS-CIC-AndMal-2020 & \makecell[c]{15} & 145 & 1.413\\
			\hline
		\end{tabular}
		\label{data-muti}
	\end{center}
	\vspace{-1.5em}
\end{table}

\subsection{Experimental evaluation of MLD-Model}
\subsubsection{Evaluation metrics}
For the problem of multi-label learning, there are currently two types of evaluation metrics. Example-based evaluation metrics (first evaluate the performance of a single sample, then average multiple samples) and label-based evaluation metrics (first consider the performance of a single label on all samples, and then average multiple labels). In this paper, we do not consider the ranking of multi-labels, so  example-based metrics are used to evaluate model.
Among these metrics, $ h $ is the model, $ N $ is the sample size, $ x_i $ is the sample, $ Y_i $ is the corresponding multi-label set, $ \Delta Y_i $ is the complement of $ Y_i $, and $ h(x_i) $ is the label set of $ x_i $ given by the model $ h $.

$Subsetacc$, evaluates the absolute accuracy of the model, i.e., the probability that all basic labels of the sample need to be successfully identified, as shown in the Eq\eqref{eqsub}.

\begin{equation}
\begin{aligned}
\operatorname{Subsetacc}&=\frac{1}{N} \sum_{i=1}^{N}|h({x}_i)=Y_i|\\ 
\end{aligned}\label{eqsub}
\end{equation}

$ Hloss $, evaluates the error of the model in detecting the multi-label of the samples, as shown in the Eq\eqref{eqhloss}. The smaller the value, the more complete the model's division ability, and the value of 0 indicates that all samples are perfectly divided.

\begin{equation}
\begin{aligned}
\operatorname{Hloss}&=\frac{1}{N} \sum_{i=1}^{N}|h({x}_i) \Delta Y_i|\\ 
\end{aligned}\label{eqhloss}
\end{equation}

The $ Precision (P)$, $ Recall (R)$, $ Accuracy (Acc)$, and $ F1$, imitate evaluation metrics in single-label classification, and are also used to calculate the comprehensive detection performance of model, as shown in the Eq\eqref{eqacc}.

\begin{equation}
	\begin{aligned}
		\operatorname{Acc}&=\frac{1}{N} \sum_{i=1}^{N} \frac{\left|Y_{i} \cap h({x}_i)\right|}{\left|Y_i \cup h\left({x}_{i}\right)\right|} \\
		\operatorname{P}&=\frac{1}{N} \sum_{i=1}^{N} \frac{\left|Y_{i} \cap h({x}_i)\right|}{|h({x}_i)|} \\
		\operatorname{R}&=\frac{1}{N} \sum_{i=1}^{N} \frac{\left|Y_{i} \cap h({x}_i)\right|}{|Y_i|} \\
		F1&=\frac{2 \times P\times R} { P+R }
	\end{aligned}\label{eqacc}
\end{equation}

\subsubsection{Environmental configuration}
The system environment is Ubuntu16.04 LTS. The hardware facilities are 16-core CPU and 128G memory. Pytorch and Scikit-learn in Python3.7 are used to implement MLD-Model. In addition, 3 NVIDIA TITAN XPs are deployed on the server.

As a rule of thumb, we set some hyper-parameters.
Samples are normalized with MinMax. The gradient penalty weight $ \lambda=10$ and our penalty item weight $ \lambda' =1$ in WGAN-GP. The unbalanced AE uses binary cross entropy to calculate the reconstruction loss.
The classifier uses cross entropy as the loss. 
The activation function is $ LeakyReLU $$ ( $negative\_slope=0.01$ ) $. 
The optimizer is $ Adam(lr$=le-3$) $.  The WGAN-GP, unbalanced AE and classifier are constructed by dense layers. The structures of MLD-Model are shown in Tab.\ref{hyper}, and the number indicates the number of neurons in the dense layer. 

\begin{table}[htbp]
	\caption{Main hyper-parameters structure of MLD-Model.}
	\vspace{-1.5em}
	\begin{center}
		\setlength{\tabcolsep}{0.9mm}{
			\begin{tabular}{|c|c|c|c|}
				\hline
				& \makecell*[c]{\textbf{WGAN-GP}} & \textbf{AE} & \textbf{Classifier}\\
				\hline
				\hline
				UNSW-NB15 & \makecell*[l]{$ G $: 100-64-128-256-42\\$ D $: 42-64-32-24-1} & \makecell*[l]{Encoder: 42-512-256-128-64\\Decoder: 64-42} & \makecell*[l]{Encoder-57}\\
				\hline
				\makecell*[l]{CCCS-CIC-\\AndMal-2020} & \makecell*[l]{$ G $: 100-128-256-512-64\\$ D $: 64-128-64-24-1} & \makecell*[l]{Encoder: 64-1024-512-256-128\\Decoder: 128-64} & \makecell*[l]{Encoder-145}\\
				\hline
		\end{tabular}}
		\label{hyper}
	\end{center}
	\vspace{-2em}
\end{table}

\subsubsection{Training set and test set}
	We perform de-duplication processing after pre-processing the network attack data with multi-label. 
	In UNSW-NB15, the official training set(after de-duplication) and test set(samples after de-duplication) are selected.
	In addition, the generated data for pre-training has 300,000 samples, with an average of 3,000 samples per single-type attack. 
	In CCCS-CIC-AndMal-2020, since there is no official training set and test set, we select to divide the dataset into training set and test set according to the ratio of 8:2.
	The generated data for pre-training has 300,000 samples, with an average of 2,000 samples per single-type attack. 
	In addition, each sample in CCCS-CIC-AndMal-2020 has 9,503-dimensional data features, so the PCA algorithm is used to reduce the dimensionality of sample to 64-dimensional after sampling.
	
	The data composition is shown in Tab.\ref{traintest}.
	The 5-fold crossover experiment is adopted in subsequent experiments. 

	\begin{table}[htbp]
		\caption{The multi-labeled data composition of training set, test set, generate data for pre-training.}
		\vspace{-1.5em}
		\begin{center}
			\begin{tabular}{|c|c|c|c|}
				\hline
				& \textbf{Trainning set} & \makecell*[c]{\textbf{Test set}} & {\textbf{Generated Data}}\\
				
				\hline
				\hline
				UNSW-NB15 & \makecell[c]{101,040} & 53,946 & 300,000\\
				CCCS-CIC-AndMal-2020 & \makecell[c]{47,550} & 11,888 & 300,000\\
				\hline
			\end{tabular}
			\label{traintest}
		\end{center}
	\end{table}

\subsection{Ablation study on key factors in MLD-Model}
\subsubsection{Data enhancement and pre-training}

\begin{figure}[htbp]
	\centerline{\includegraphics[scale=0.4]{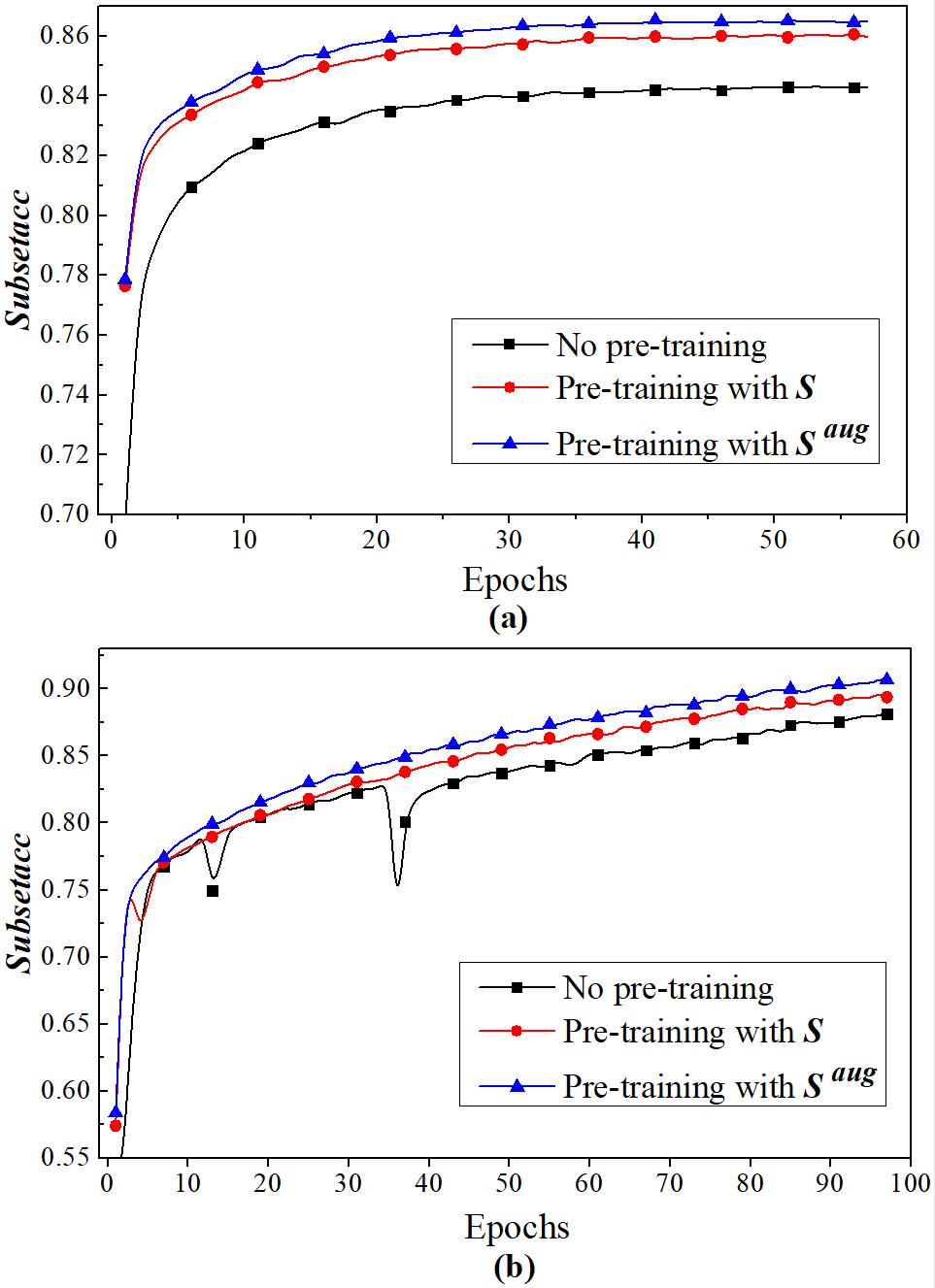}}
	\caption{The $ Subsetacc $ convergence results of MLD-Model in the classifier phase in training set. (a) In UNSW-NB15, (b) In CCCS-CIC-AndMal-2020.}
	\label{pretaintu}
\end{figure}

We believe that data enhancement and pre-training can effectively enable MLD-Model to obtain better detection performance. Fig.\ref{pretaintu} shows the convergence results of training $ Subsetacc $ of the classifier in MLD-Model in different conditions.
In UNSW-NB15 (Fig.\ref{pretaintu}(a)), the classifier can reach convergence in the three conditions. The classifier based on pre-training with $ S^{aug} $ can obviously converge faster and better. 
In CCCS-CIC-AndMal-2020, the classifier based on no pre-training has large fluctuations in the process of convergence. 
And the classifier of MLD-Model based on pre-training with $ S^{aug} $ get the best convergence effect.
In general, the classifier can get better convergence results after pre-training. Especially, the classifier can further converge faster and higher after adding the generated data.

In actual detection, MLD-Model with pre-training, especially adding the generated data, also has a better detection results. Tab.\ref{yuxun1} shows the detection results of MLD-Model in different conditions in UNSW-NB15 and CCCS-CIC-AndMal-2020. The classifier of MLD-Model based on pre-training with $ S^{aug} $ can get the best overall results. For instance, in UNSW-NB15, the $ Hloss $ drops from 0.446 to 0.425, and the $ F1 $ increases from 78.72\% to 80.06\%. In CCCS-CIC-AndMal-2020, the $ Hloss $ drops from 0.365 to 0.342, and the $ F1 $ increases from 82.49\% to 83.63\%. 

\begin{table}[htbp]
	\caption{The detection results of MLD-Model in UNSW-NB15}
	\vspace{-1.5em}
	\begin{center}
		\setlength{\tabcolsep}{0.7mm}{
			\begin{tabular}{|l|c|c|c|c|c|c|}
				\hline
				& \makecell*[l]{$ \mathbf{Subsetacc} $} & $ \mathbf{Hloss} $ & $ \mathbf{Acc} $ & $ \mathbf{P} $ & $ \mathbf{R} $ & $ \mathbf{F1} $\\
				\hline				
				\hline
				\multicolumn{7}{|l|}{\makecell*[l]{\textbf{In UNSW-NB15}}} \\
				\hline
				\makecell[l]{No pre-training} & 78.01\% & 0.446 & 78.57\% & 78.71\% & 78.74\% & 78.72\% \\ 
				\makecell[l]{Pre-training with $ S $} & 78.43\% & 0.447 & 78.90\% & 79.12\% & 79.06\% & 79.09\% \\ 
				\makecell[l]{Pre-training with $ S^{aug} $} & \textbf{79.27\%} & \textbf{0.425} & \textbf{79.87\%} & \textbf{80.03\%} & \textbf{80.09\%} & \textbf{80.06\%} \\ 
				\hline
				\multicolumn{7}{|l|}{\makecell*[l]{\textbf{In CCCS-CIC-AndMal-2020}}} \\
				\hline
				\makecell[l]{No pre-training} & 81.31\% & 0.365 & 82.08\% & 82.73\% & 82.25\% & 82.49\% \\ 
				\makecell[l]{Pre-training with $ S $} & 81.53\% & 0.359 & 82.40\% & 83.08\% & 82.67\% & 82.87\% \\ 
				\makecell[l]{Pre-training with $ S^{aug} $} & \textbf{82.28\%} & \textbf{0.342} & \textbf{83.17\%} & \textbf{83.83\%} & \textbf{83.43\%} & \textbf{83.63\%} \\ 
				\hline
		\end{tabular}}
		\label{yuxun1}
	\end{center}
    \vspace{-0.5em}
\end{table}

In addition, we also try to generate data using WGAN-GP with traditional loss(Eq\eqref{eq1}), which is then used for pre-training. However, the result is fluctuating, sometimes making the model even worse than without pre-training. Therefore, we finally select WGAN-GP with improved loss(Eq\eqref{eq2}) to generate data.

\subsubsection{Time complexity}
The performance of MLD-Model in UNSW-NB15 and CCCS-CIC-AndMal-2020 is similar. We compare the time required for MLD-Model to achieve $ Subsetacc $=84\% in training set under three conditions in UNSW-NB15 and to achieve $ Subsetacc $=88\% in training set under three conditions in CCCS-CIC-AndMal-2020. The results are shown in Tab.\ref{yuxunxxx1}.
Since the data generation belongs to preliminary preparation, we ignore this part of the time. Tab.\ref{yuxunxxx1} shows that pre-training can reduce the model convergence time in the fine-tuning phase.
Although pre-training with $ S^{aug} $ increases the pre-training time, it can further reduce the fine-tuning time.

\begin{table}[htbp]
	\caption{The training and detection time of MLD-Model in UNSW-NB15}
	\vspace{-1.5em}
	\begin{center}
		\setlength{\tabcolsep}{0.8mm}{
			\begin{tabular}{|l|c|c|c||c|}
				\hline
				& \makecell*[l]{\textbf{Pre-training}} & \textbf{Fine-tuning} & \textbf{Detection} & \textbf{Total}\\
				\hline
				\hline
				\multicolumn{5}{|l|}{\makecell*[l]{\textbf{In UNSW-NB15}}} \\
				\hline
				\makecell[l]{No pre-training} & \textbf{0s} & 97.40s & \textbf{3.22s} & 100.62s \\ 
				\makecell[l]{Pre-training with $ S $} & 9s & 31.28s & 3.41s & \textbf{43.69s} \\ 
				\makecell[l]{Pre-training with $ S^{aug} $} & 18.9s & \textbf{22.97s} & 3.34s & 45.21s \\
				\hline
				\multicolumn{5}{|l|}{\makecell*[l]{\textbf{In CCCS-CIC-AndMal-2020}}} \\
				\hline
				\makecell[l]{No pre-training} & \textbf{0s} & 119.01s & 0.93s & 119.94s \\ 
				\makecell[l]{Pre-training with $ S $} & 12.6s & 92.97s & 0.93s & \textbf{106.50s} \\ 
				\makecell[l]{Pre-training with $ S^{aug} $} & 30.80s & \textbf{76.86s} & 0.93s & 108.59s \\ 
				\hline
		\end{tabular}}
		\label{yuxunxxx1}
	\end{center}
\end{table}

\subsection{Comparison with other methods}

There is no related multi-label methods for detecting the network attacks in the overlapping phenomenon.
Therefore, we select several baseline methods for comparison. 

The methods selected are shown in Tab.\ref{alg}. Based on the 3 strategies(first-order, second-order, high-order), algorithms can be divided into problem transformation and algorithm adaptation.
 In the problem transformation, we select 3 transformation strategies: Binary Relevance(BR)\cite{boutell2004learning}, Calibrated Label Ranking(CLR)\cite{furnkranz2008multilabel}, and Classifier Chains(CC)\cite{read2009classifier, read2011classifier}. In the algorithm adaptation, we select the ML-KNN algorithm\cite{zhang2007ml} that belongs to lazy learning\cite{aha2013lazy}. 
	
\begin{table}[htbp]
	\caption{Multi-label learning methods selected for comparison.}
	\begin{center}
		\begin{tabular}{|c|c|c|c|}
			\hline
			\makecell*[c]{\textbf{Method}} & \makecell*[c]{\textbf{Type}} & \textbf{Strategy} & \textbf{Re-name}\\
			\hline
			\hline
			\makecell[c]{Bayes} & {BR} & {first-order} & {Bayes-BR} \\
			\makecell[c]{Logistic Regression} & {BR} & {first-order} & {LR-BR} \\
			\makecell[c]{Decision Tree} & {BR} & {first-order} & {DT-BR} \\
			\makecell[c]{Random Forest} & {BR} & {first-order} & {RF-BR} \\
			\makecell[c]{SVM} & {BR} & {first-order} & {SVM-BR} \\
			\hline
			\makecell[c]{Bayes} & {CLR} & {second-order} & {Bayes-CLR} \\
			\makecell[c]{Logistic Regression} & {CLR} & {second-order} & {LR-CLR} \\
			\makecell[c]{Decision Tree} & {CLR} & {second-order} & {DT-CLR} \\
			\makecell[c]{Random Forest} & {CLR} & {second-order} & {RF-CLR} \\
			\makecell[c]{SVM} & {CLR} & {second-order} & {SVM-CLR} \\
			\hline
			\makecell[c]{Bayes} & {CC} & {high-order} & {Bayes-CC} \\
			\makecell[c]{Logistic Regression} & {CC} & {high-order} & {LR-CC} \\
			\makecell[c]{Decision Tree} & {CC} & {high-order} & {DT-CC} \\
			\makecell[c]{Random Forest} & {CC} & {high-order} & {RF-CC} \\
			\makecell[c]{SVM} & {CC} & {high-order} & {SVM-CC} \\
			\hline
			\makecell[c]{ML-KNN\cite{zhang2007ml}} & \makecell[c]{Lazy Learning} & {first-order} & ML-KNN \\ 
			\hline
		\end{tabular}
		\label{alg}
	\end{center}
\end{table}

\subsubsection{UNSW-NB15}
In UNSW-NB15, as shown in Tab.\ref{nbqita}, MLD-Model is superior to other methods in 4 of the 5 metrics.
The Bayes-BR can get the best results on recall $ R $=99.74\%, but it fails in other metrics, such as the $ F1 $ of 43.14\% and the $ Hloss $ of 5.227. This means that Bayes-BR considers that a sample belongs to almost all categories when it detects a sample, which is unreasonable.
In general, MLD-Model can get the better overall performance compared with other methods.

		\begin{table}[htbp]
		\caption{Detection results of MLD-Model and other methods in UNSW-NB15}
		\vspace{-1.5em}
		\begin{center}
			\setlength{\tabcolsep}{0.9mm}{
				\begin{tabular}{|c|c|c|c|c|c|c|}
					\hline
					\textbf{Method} & \makecell*[c]{$ \mathbf{Subsetacc} $} & $ \mathbf{Hloss} $ & $ \mathbf{Acc} $ & $ \mathbf{P} $ & $ \mathbf{R} $ & $ \mathbf{F1} $\\
					\hline
					\hline
					\makecell[c]{Bayes-BR} & {10.60\%} & {5.227} & {27.48\%} & {27.52\%} & {\textbf{99.74\%}} & {43.14\%} \\
					\makecell[c]{LR-BR} & {50.72\%} & {0.635} & {51.73\%} & {51.88\%} & {52.55\%} & {52.21\%} \\
					\makecell[c]{DT-BR} & {65.15\%} & {0.543} & {70.01\%} & {70.08\%} & {74.8\%} & {72.36\%} \\
					\makecell[c]{RF-BR} & {72.36\%} & {0.450} & {73.41\%} & {73.47\%} & {74.09\%} & {73.78\%} \\
					\makecell[c]{SVM-BR} & {48.81\%} & {0.603} & {49.13\%} & {49.17\%} & {49.24\%} & {49.2\%} \\
					\hline
					\makecell[c]{Bayes-CLR} & {23.66\%} & {1.765} & {33.02\%} & {33.28\%} & {42.22\%} & {37.22\%} \\
					\makecell[c]{LR-CLR} & {63.83\%} & {0.736} & {64.24\%} & {64.64\%} & {64.33\%} & {64.49\%} \\
					\makecell[c]{DT-CLR} & {73.65\%} & {0.539} & {74.25\%} & {74.35\%} & {74.52\%} & {74.43\%} \\
					\makecell[c]{RF-CLR} & {76.61\%} & {0.478} & {77.19\%} & {77.24\%} & {77.41\%} & {77.33\%} \\
					\makecell[c]{SVM-CLR} & {63.85\%} & {0.734} & {64.3\%} & {64.7\%} & {64.42\%} & {64.56\%} \\
					\hline
					\makecell[c]{Bayes-CC} & {8.82\%} & {3.662} & {15.85\%} & {15.99\%} & {34.26\%} & {21.81\%} \\
					\makecell[c]{LR-CC} & {68.85\%} & {0.644} & {69.21\%} & {69.47\%} & {69.35\%} & {69.41\%} \\
					\makecell[c]{DT-CC} & {73.09\%} & {0.541} & {74.22\%} & {74.31\%} & {75.08\%} & {74.69\%} \\
					\makecell[c]{RF-CC} & {76.94\%} & {0.451} & {77.54\%} & {77.59\%} & {77.77\%} & {77.68\%} \\
					\makecell[c]{SVM-CC} & {68.75\%} & {0.642} & {69.24\%} & {69.40\%} & {69.48\%} & {69.44\%} \\
					\hline
					\makecell[c]{ML-KNN} & 64.97\% & 0.590 & 65.78\% & 65.94\% & 66.24\% & 66.09\% \\
					\hline
					\makecell[c]{MLD-Model} & \textbf{79.27\%} & \textbf{0.425} & \textbf{79.87\%} & \textbf{80.03\%} & 80.09\% & \textbf{80.06\%} \\
					\hline
			\end{tabular}}
			\label{nbqita}
		\end{center}
	\end{table}

\begin{table}[htbp]
	\caption{Detection results of MLD-Model and other methods in CCCS-CIC-AndMal-2020}
	\vspace{-1.5em}
	\begin{center}
		\setlength{\tabcolsep}{0.9mm}{
			\begin{tabular}{|c|c|c|c|c|c|c|}
				\hline
				\textbf{Method} & \makecell*[c]{$ \mathbf{Subsetacc} $} & $ \mathbf{Hloss} $ & $ \mathbf{Acc} $ & $ \mathbf{P} $ & $ \mathbf{R} $ & $ \mathbf{F1} $\\
				\hline
				\hline
				\makecell[c]{Bayes-BR} & {22.8\%} & {3.715} & {39.19\%} & {39.25\%} & {\textbf{91.46\%}} & {54.93\%} \\
				\makecell[c]{LR-BR} & {61.8\%} & {0.458} & {62.08\%} & {62.37\%} & {62.11\%} & {62.24\%} \\
				\makecell[c]{DT-BR} & {67.07\%} & {0.546} & {71.65\%} & {72.2\%} & {76.39\%} & {74.24\%} \\
				\makecell[c]{RF-BR} & {68.76\%} & {0.357} & {69.76\%} & {70.35\%} & {70.2\%} & {70.28\%} \\
				\makecell[c]{SVM-BR} & {59.62\%} & {0.47} & {59.76\%} & {59.92\%} & {59.76\%} & {59.84\%} \\
				\hline
				\makecell[c]{Bayes-CLR} & {52.83\%} & {0.920} & {58.97\%} & {59.2\%} & {64.98\%} & {61.96\%} \\
				\makecell[c]{LR-CLR} & {73.81\%} & {0.521} & {74.54\%} & {75.31\%} & {74.54\%} & {74.92\%} \\
				\makecell[c]{DT-CLR} & {77.37\%} & {0.439} & {78.45\%} & {79.22\%} & {78.8\%} & {79.01\%} \\
				\makecell[c]{RF-CLR} & {80.47\%} & {0.378} & {81.46\%} & {82.26\%} & {81.69\%} & {81.98\%} \\
				\makecell[c]{SVM-CLR} & {74.3\%} & {0.510} & {75.06\%} & {75.86\%} & {75.06\%} & {75.46\%} \\ 
				\hline
				\makecell[c]{Bayes-CC} & {54.76\%} & {1.331} & {58.13\%} & {58.28\%} & {65.21\%} & {61.55\%} \\
				\makecell[c]{LR-CC} & {70.35\%} & {0.588} & {71.06\%} & {71.83\%} & {71.06\%} & {71.44\%} \\
				\makecell[c]{DT-CC} & {72.37\%} & {0.555} & {74.27\%} & {74.94\%} & {75.75\%} & {75.34\%} \\
				\makecell[c]{RF-CC} & {73.42\%} & {0.334} & {74.45\%} & {75.1\%} & {74.87\%} & {74.99\%} \\
				\makecell[c]{SVM-CC} & {68.95\%} & {0.615} & {69.69\%} & {70.5\%} & {69.69\%} & {70.09\%} \\
				\hline
				\makecell[c]{ML-KNN} & 72.59\% & 0.382 & 73.48\% & 73.94\% & 73.92\% & 73.93\% \\
				\hline
				\makecell[c]{MLD-Model} & \textbf{82.28\%} & \textbf{0.342} & \textbf{83.17\%} & \textbf{83.83\%} & 83.43\% & \textbf{83.63\%} \\ 
				\hline
		\end{tabular}}
		\label{qita202}
	\end{center}
	\vspace{-1.0em}
\end{table}

\subsubsection{CCCS-CIC-AndMal-2020}
In CCCS-CIC-AndMal-2020, as shown in Tab.\ref{qita202}, MLD-Model is also superior to other methods in 4 of the 5 metrics.
Similarly, although the Bayes-BR can get the best results on recall $ R $=91.46\%, it fails in other metrics, such as the $ F1 $ of 54.93\% and the $ Hloss $ of 3.715.
This means that Bayes-BR considers that a sample belongs to almost all categories when it detects a sample, which is unreasonable.
Our method can get the $ F1 $ of 83.63\% and the $ Hloss $ of 0.342 in CCCS-CIC-AndMal-2020. In general, MLD-Model can get the better overall performance compared with other methods.

\section{Discussion}

\subsection{Hyper-parameter selection}
	MLD-Model has a large number of hyper-parameters, such as the number of layers in the neural network, the learning rate, the epoches, and batch-size. In the experiment, we make selections based on experience and previous experimental basis. For multiple optional hyper-parameters, grid search is used to select the optimal hyper-parameter combination.

\subsection{Comparison with related single-label methods}

The sample is multi-labeled in the overlapping phenomenon of behavior attribute, and one sample could belong to multiple attacks.
However, traditional single-label methods are difficult to detect multi-label network attacks.
The single-label network attack detection methods can only give a single-category for a sample, which means that the theoretical upper limit of the accuracy is less than 100\%.
They perform poorly when applied to multi-label network attack detection.

Taking the mutlti-label network attack detection in UNSW-NB15 as an example, we implement the ICVAE-DNN\cite{yang2019improving} and SAVAER-DNN\cite{yang2020network}. The results are shown in Tab.\ref{qitabi}. The performance of these two methods is lower than MLD-Model because only one category can be given for each sample. Therefore, MLD-Model is more suitable for network attack detection in the overlapping phenomenon than single-label detection methods.

\begin{table}[htbp]
	\caption{Detection results of MLD-Model and related single-label methods in UNSW-NB15}
	\vspace{-1.5em}
	\begin{center}
		\setlength{\tabcolsep}{0.7mm}{
			\begin{tabular}{|c|c|c|c|c|c|c|}
				\hline
				\textbf{Method} & \makecell*[c]{$ \mathbf{Subsetacc} $} & $ \mathbf{Hloss} $ & $ \mathbf{Acc} $ & $ \mathbf{P} $ & $ \mathbf{R} $ & $ \mathbf{F1} $\\
				\hline
				\hline
				\makecell*[c]{ICVAE-DNN} & {71.55\%} & {0.589} & {71.76\%} & {72.42\%} & {71.76\%} & {72.09\%} \\
				\makecell*[c]{SAVAER-DNN} & {73.43\%} & {0.548} & {73.68\%} & {74.47\%} & {73.68\%} & {74.07\%} \\
				\hline
				\makecell*[c]{MLD-Model} & \textbf{79.27\%} & \textbf{0.425} & \textbf{79.87\%} & \textbf{80.03\%} & \textbf{80.09\%} & \textbf{80.06\%} \\
				\hline
		\end{tabular}}
		\label{qitabi}
	\end{center}
	\vspace{-1.5em}
\end{table}


\subsection{Performance of MLD-Model} 
Although the overall performance of MLD-Model is better than other methods, its various metrics are only 80+\% ($ Acc $, $ F1 $, \textit{etc.}). There are two main reasons, the increased output space and the complexity of the network attack itself.

First, a single-label network attack is multi-labeled in the overlapping phenomenon of behavior attribute. 
In UNSW-NB15, the output space is increased from 10 categories to 57 categories after data processing and analysis.
The increase in output space make detection more difficult. 

Second, there is not only overlapping phenomenon of behavior attribute between network attacks, but also similar phenomenon.
Similar phenomenon represents that attacks are not completely consistent, but very similar. The difference between two samples is less than a threshold $ \epsilon $. This will make they difficult to be distinguished, resulting in false positives and false negatives of model. For instance, as shown in Fig.\ref{abcdef} in Section 3, except for overlapping phenomenon, there is also a similar phenomenon of behavior attribute between network attack A and B.
The research about similar phenomenon of behavior attribute is also our future work.

\subsection{Application scenario of MLD-Model} 

The overlapping phenomenon cause that a sample could be multi-labeled. And if a sample is multi-labeled, traditional single-label detection methods only can only give one label to this sample at most, which will lead to false negatives, making its theoretical accuracy impossible to be 100\%. This fundamentally limits the effectiveness of traditional methods. 
MLD-Model uses a multi-label approach to detect network attacks in overlapping phenomenon, which can guarantee the theoretical line of 100\% accuracy. More importantly, MLD-Model can help network administrators in two aspects. 

\textbf{Tracing the source of network attack:} A sample detected as multi-label can reflect more information about the attacker behind it, such as more detailed attack methods, so that we can find the attacker more easily.

\textbf{Building a better IDS:} In a specific scenario, the distribution and similarity between different attacks can be obtained by analyzing and detecting the network attacks in the overlapping phenomenon in a multi-label manner, then helping us to build a more comprehensive defense scheme.

\section{Conclusion}
In this paper, we discover the overlapping phenomenon of behavior attribute between network attacks in the real world, and analyze the reasons. Experiments also verifies our conclusions. 
Overlapping phenomenon leads to a network attack sample may be multi-labeled. Identifying these attacks with a multi-label method can help researchers better trace the source of network attacks and build a better IDS.
Therefore, we also propose a multi-label detection method, MLD-Model, in which WGAN-GP with improved loss is used for data enhancement, and unbalanced AE is used for pre-training. Finally, MLD-Model can achieve $ F1 $=80.06\% in UNSW-NB15 and $ F1 $=83.63\% in CCCS-CIC-AndMal-2020.

In the future, we will further explore the correlation between network attacks, 
and explore more suitable detection methods.

\section*{Acknowledgements}
This work is supported by the National Key Research and Development Program of China\\ (Grant No.2018YFB0804704), and the National Key Research and Development Program of \\China (Grant No.2019YFB1005201). 

\bibliographystyle{cas-model2-names}

\bibliography{Bibliography}



\end{document}